# Low Temperature Thermal Transport in Partially Perforated Silicon Nitride Membranes

V. Yefremenko, G. Wang, V. Novosad [(*)], A. Datesman, and J. Pearson

*Materials Science Division, Argonne National Laboratory, IL 60439*

R. Divan

*Center for Nanoscale Materials, Argonne National Laboratory, IL 60439*

C. Chang, T. Downes, J. McMahon, L. Bleem, A. Crites, S. S. Meyer, and J. E. Carlstrom

*Kavli Institute for Cosmological Physics, University of Chicago, Chicago, IL 60637*

**ABSTRACT**

The thermal transport in partially trenched silicon-nitride membranes has been studied in the temperature range from 0.3 to 0.6 K, with the Transition Edge Sensor (TES) the sole source of membrane heating. The test configuration consisted of Mo/Au TESs lithographically defined on silicon nitride membranes 1 micron thick and 6 millimeters square in size. Trenches with variable depth were incorporated between the TES and the silicon frame in order to manage the thermal transport. It was shown that sharp features in the membrane surface, such as trenches, significantly impede the modes of phonon transport. A non-linear dependence of thermal resistance on trench depth was observed. Partial perforation of silicon nitride membranes to control thermal transport could be useful in fabricating mechanically robust detector devices.

(*) Corresponding author. Electronic address: *novosad@anl.gov*

Thermal transport in silicon nitride membranes defines the basic parameters of a variety of devices operating at low temperatures. Especially important is the thermal link characterization of Transition Edge Sensors (TES), which represent an attractive bolometric technology for broadband detectors [1,2]. Because thermal transport determines both the dynamic and static operation of bolometric detectors, control of the thermal conductance to bath, *G*, is critical for the implementation of detectors utilizing TES. Theoretical [3] - [6] and experimental [7] - [10] investigations of thermal transport in both continuous and perforated or bridge supported silicon nitride membranes have been conducted by researchers in numerous laboratories. In dielectric solids, heat is carried by phonons and the thermal conductivity $\kappa$ is proportional to the phonon mean free path. For amorphous solids below 1 K, the mean free path is roughly on the scale of microns and varies approximately as $T^{-1.2}$ [11]. As the thickness of the membrane becomes smaller than the phonon mean free path in bulk material, the interaction of phonons with the surface should play a significant role in the scattering mechanism. Phonon scattering by surface asperities was considered in [13], [17] where the specular reflection coefficient *f (λ)* was estimated from the root-mean square surface roughness $\eta$ and the dominant phonon wavelength $\lambda_{dom}$:

$$f(\lambda) = exp(-16 \pi^3 \eta^2 / \lambda_{dom}^2) \tag{1}$$

The phonon wavelength $\lambda_{dom}$ increases with decreasing temperature according to [11]:

$$\lambda_{dom} = 0.235 h c_s / kT, \tag{2}$$

where h and k are Plank and Boltzmann's constant respectively, $c_s$ speed of sound. Thus phonon reflection from the surface is very sensitive to the surface roughness. The ballistic regime resulting from specular surface scattering was observed in silicon nitride membranes in the temperature range <0.2K [8], [10]. It has been shown that additional scattering may be introduced by even very small surface inhomogeneities such as thin films [12], [13] or small particles [8]. Completely diffusive surface scattering should occur when the dimensions of surface defects or of artificial features are of the order of the mean free path [8], [13].

We have recently reported measurements of the thermal conductance, G, for a TES detector on a 3 × 3 mm$^2$ silicon nitride membrane utilizing a partially perforated trench, of 800 pW/K. This compares to 3nW/K which was measured for a device on an un-patterned membrane lacking thermal control structures [14]. Based upon this understanding, we have examined the use of partially perforated membranes for thermal transport management in sub-Kelvin temperature range. Our experiments examined the thermal transport in absorber coupled TES bolometers on partially perforated Si-N membranes [15]. The tested experimental structure is shown in Fig. 1.

For sample preparation we utilized commercial amorphous low-stress Si-N films 1 μm thick grown by low pressure chemical vapor deposition on to (100) Si wafers [16]. The 6 mm x 6 mm Si-N membrane window was fabricated using standard anisotropic KOH etching of Si. Two rows of trenches, each 20 μm wide and ~3 mm in diameter, were fabricated using CF$_4$ plasma etching. The distance between the two rows of trenches was 60 μm. The depth of the trenches varied from zero (a continuous membrane) to the full depth of 1 μm (a through-etched membrane). Device fabrication began with DC magnetron sputter deposition of a bilayer film of 50nm Mo/50nm Au on to the surface of a Si/Si-N wafer at room temperature. Deposition was performed under a single vacuum following RF cleaning of the substrate. The 50 μm × 50 μm TES was subsequently patterned using optical lithography and wet etching. We optimized the transition temperature of our TES thermometers in the range of 400 – 600 mK utilizing the superconductor-normal metal proximity effect. For the devices just described, we found $T_C$ = 530 ± 10 mK, δT < 10 mK, and $R_n$ = 0.3 Ω. Each device also contains absorber elements deposited on the central portion of the membrane. Additional structures on the surface might influence the phonon scattering, but since they are essentially identical for all tested devices we neglect their role in thermal transport through the trenched portion of the membrane.

The TES leads consist of superconducting Nb, with a total length between the sensor and the Si frame of 3.83 mm. Various widths (5 – 20 µm) and thicknesses (100 – 150 nm) were employed for the leads. Deposition and lift-off processes were used for both absorbers and leads fabrication. Experiments were conducted using a $He^3$ cryostat in the temperature range from 0.3 to 0.6 K, with the TES detector as the sole source of membrane heating. In this case, the power flow from the TES at temperature $T$ to the bath at temperature $T_{bath}$ can be expressed as

$$P=K(T^n-T^n_{bath}), \quad (3)$$

which implies a thermal conductance

$$G=nKT^{n-1} \quad (4)$$

Here, $K$ and $n$ are coefficients which characterize the thermal conductance and depend upon both geometry and material.

The measured TES resistance dependence as a function of $T_{bath}$, measured for a range of bias currents, was used to calculate the thermal conductance $G=dP/dT$, $K$, and $n$ by fitting to the curve of dissipated power versus bath temperature. The TES operating temperature $T$ decreases only slightly from the zero current value of $T_C$ due to the device current. We estimate TES critical current at 0 K is approximately $I_{Co}=5mA$. Thus in the worst case, with the largest bias current ~ 50µA at the lowest bath temperature, $T_c$ decreases by less than 5%. The thermal conductance $G=nKT^{n-1}$ as a function of base temperature for different devices is summarized in Fig. 2. Trenched samples exhibit identical values of the index $n$, while the values for continuous and through-etched membranes are completely different. The earlier reported [7] values of $n=3.05$ and $n=2.98$ for a continuous membrane and $n=2.54$ for a bridge suspended membrane are in general agreement with these results. Because of the dependence of the thermal conductivity upon the mean free path, therefore, the changing values for $n$ indicated in Fig. 2 are evidence that an additional scattering mechanism occurs in trenched membranes which is not present in a continuous membrane.

The results for trenched and continuous membranes lie below $n=4$ typical for purely ballistic transport and/or completely diffusive surface scattering. In these two model cases the probability of specular reflection, as well as the phonon mean free path are temperature independent. In a real sample the strength of the phonon mean free path temperature dependence is function of the ratio between surface roughness and dominant phonon wavelength. With lowering temperature the average phonon wavelength increases and consequently a surface of given roughness appears smoother. A stronger temperature dependence of the phonon mean free path leads to more significant changes of index $n$. This is consistent with the interpretation that the higher value of the index $n$ for the continuous membrane reflects a dominant ballistic contribution to the thermal conductance, which is significantly reduced in trenched samples. Identical values of $n$ for the samples with variable trench depths indicate that the trench design (width, spacing, and number of trenches), rather than its depth, activates an additional scattering mechanism.

Our result for a completely perforated membrane, $n=2.54$, agrees exactly with a result published by another group [7] for a suspended membrane. This is explained [5] as a result of the interaction of phonons with the edges of the sample.

With decreasing values of $n$, our data indicates stronger temperature dependence for the phonon mean free path in the trenched devices in comparison with the continuous membrane.

The thermal transport contribution of Nb leads was separately estimated using test samples consisting of a TES together with a heater on narrow Si-N bridges of various lengths. In this experimental configuration, the thermal conductivity of the leads $G_{leads}$ was large compared to the parallel thermal conductance of the membrane. Using the data for electron-phonon coupling strength and Kapitza thermal conductivity from [10, 18] we find the electron–phonon thermal conductance $G_{ep}= 3.9 \times 10^{-8}$ W/K and Kapitza thermal conductance between TES and membrane $G_{kap}= 4.6 \times 10^{-8}$

W/K . Utilizing this data and the known geometries of the samples, we estimate the contribution of the leads is negligible (<5%) compared to the measured value for the thermal conductance.

The thermal conductance as a function of trench depth for T~0.5 K is shown in Fig.3. The surface roughness $\eta$ and the depth of trenches was separately evaluated using an atomic force microscope (see insert in Fig. 3). For measured roughness $\eta = 11.4$ nm the corresponding values of the specular reflection coefficients from equation (1), are about 1% and ~40% for the temperatures 0.5K and 0.3K respectively. Thus, the additional thermal resistance includes phonon reflection and scattering from the quasi-vertical trench walls, and surface scattering in the trench bottom. The total thermal conductance $G_o$ can be considered as a serial connection of the trenched and continuous portions of the membrane $1/G_o = 1/G_{tr} + 1/G_c$. The temperature distribution is not uniform across the trenches and therefore it is difficult to separately extract the index $n$ for the trenches and surrounding membrane. However, noting that when the bath temperature is close to the $T_c$ the membrane temperature distribution can be considered quasi-isothermal, and using the value of $G_c$ for a continuous membrane to represent the thermal conductance of the continuous portions of trenched membranes, it is possible to extract values of $G_{tr}$ (Fig. 3). Deeper trenches lead to higher thermal resistance. We expect this to be the case up until the point at which the residual thickness $b < h\nu/2kT$ (~50 nm). At this point, the crossover from 3D-2D will lead to an increase in the thermal conductance [5]. A non-linear dependence of thermal resistance on trench depth was observed.

To summarize, sharp features in the membrane surface, such as trenches, significantly impede the modes of phonon transport. Our measurements have demonstrated that partial perforation is an effective technique for thermal transport control at low temperatures. Trenched membranes are more robust mechanically than perforated membranes. This advantage will increase the feasibility and reliability of large arrays of high sensitivity superconducting TES detectors.

The authors greatly acknowledge NIST Quantum Devices Group at Boulder for providing SQUID array for TES readout. We thank Matthew E. Kenyon from JPL for useful discussions of thermal transport and device microfabrication aspects. The work at Argonne National Laboratory, including the use of the facilities at the Center for Nanoscale Materials (CNM), was supported by UChicago Argonne, LLC, Operator of Argonne National Laboratory ("Argonne"). Argonne, a U.S. Department of Energy Office of Science Laboratory, is operated under Contract No. DE-AC02-06CH11357.

**Figure Captions:**

Fig.1 Optical microscope image of an absorber coupled TES bolometer.

Fig. 2 Thermal conductance for samples with different depth of trenches vs. bath temperature.

Fig. 3 Summarized dependence of thermal conductance $G_0$ for all tested samples at 0.5K. The insert shows separately estimated $G_{tr}$ at this temperature, and an Atomic Force Microscope image of the trenched portion of membrane.

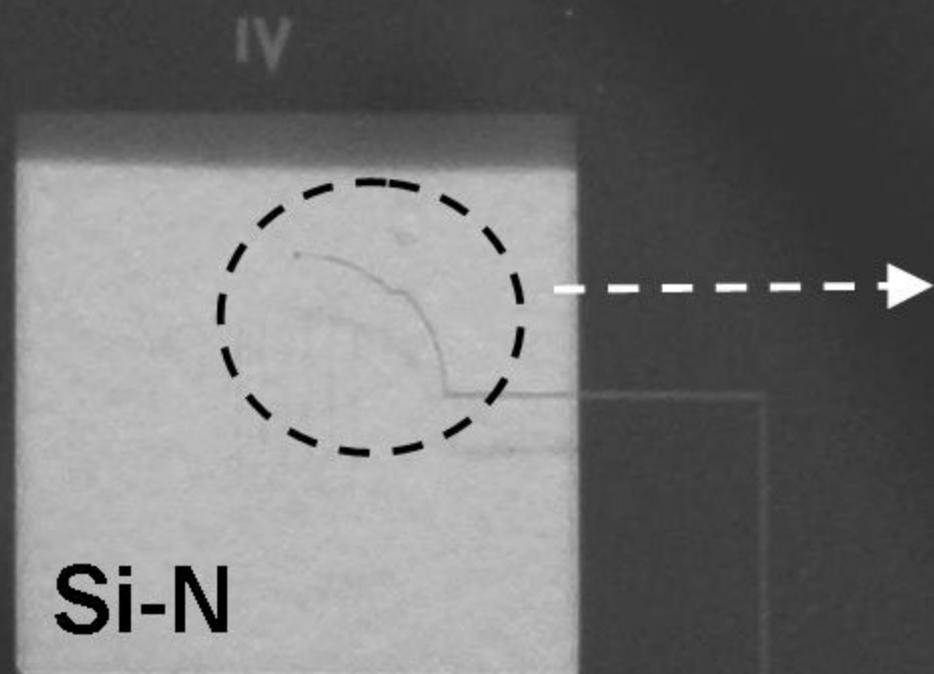

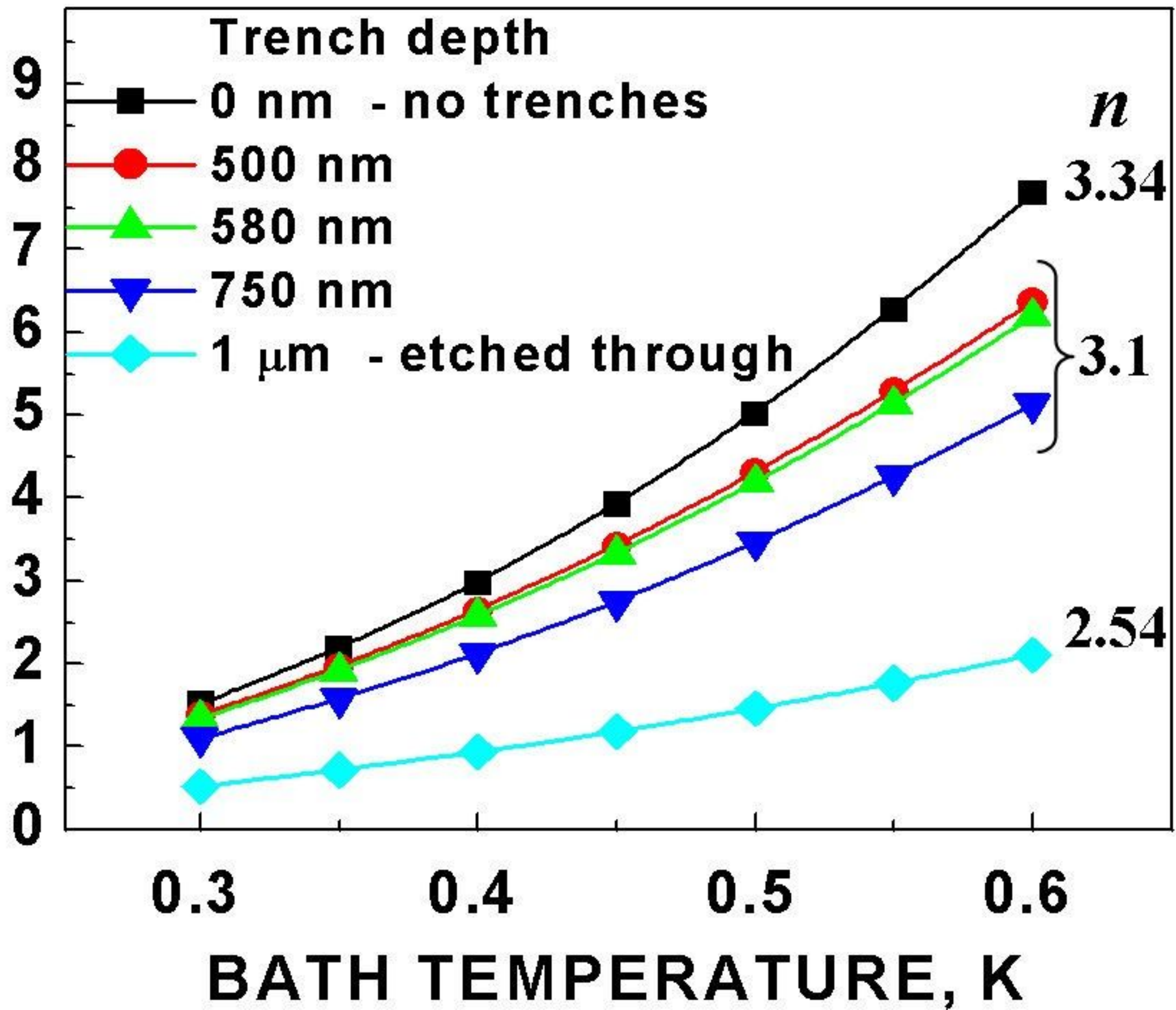

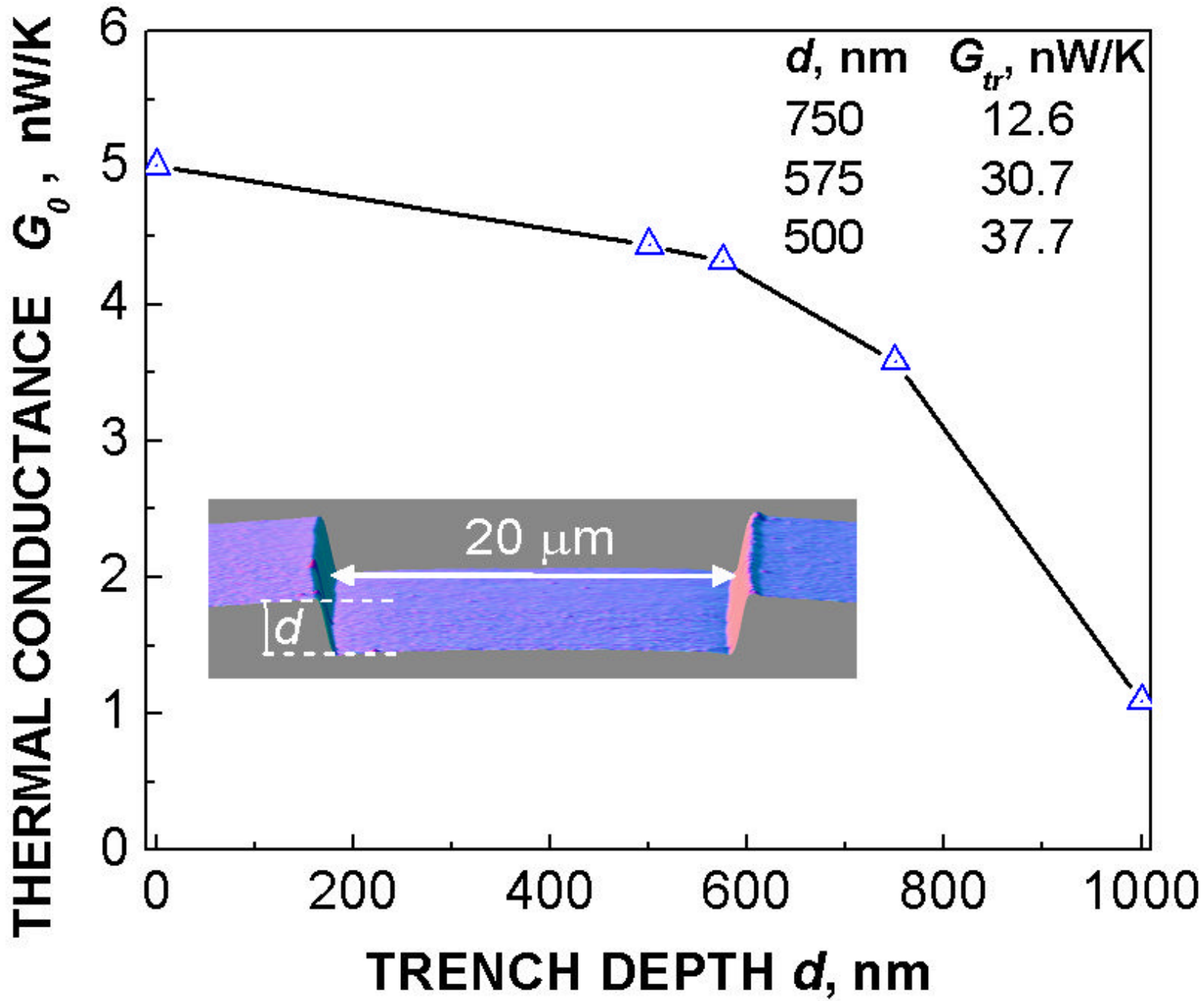

| $d$, nm | $G_{tr}$, nW/K |
|---|---|
| 750 | 12.6 |
| 575 | 30.7 |
| 500 | 37.7 |